\documentclass[Physsubmission, Phys]{SciPost}

\hypersetup{
    colorlinks,
    linkcolor={red!50!black},
    citecolor={blue!50!black},
    urlcolor={blue!80!black}
}

\usepackage[bitstream-charter]{mathdesign}
\renewcommand\slash[1]{\not \! #1}

\DeclareSymbolFont{usualmathcal}{OMS}{cmsy}{m}{n}
\DeclareSymbolFontAlphabet{\mathcal}{usualmathcal}

\begin{document}

\begin{center}{\Large \textbf{
The sensitivity to BSM in di-taon production at the LHC\\
}}\end{center}

\begin{center}
Mariola K{\l}usek-Gawenda\textsuperscript{1$\star$},
Antoni Szczurek\textsuperscript{1,2},
Mateusz Dynda{\l}\textsuperscript{3} and
Matthias Schott\textsuperscript{4}
\end{center}

\begin{center}
{\bf 1} Institute of Nuclear Physics Polish Academy of Sciences, PL-31342 Krakow, Poland
\\
{\bf 2} Faculty of Mathematics and Natural Sciences, University of Rzeszow, Poland
\\
{\bf 3} AGH University of Science and Technology, Krakow, Poland 
\\
{\bf 4} Johannes Gutenberg University, Mainz, Germany\\
* mariola.klusek@ifj.edu.pl
\end{center}

\begin{center}
\today
\end{center}


\definecolor{palegray}{gray}{0.95}
\begin{center}
\colorbox{palegray}{
  \begin{tabular}{rr}
  \begin{minipage}{0.1\textwidth}
    \includegraphics[width=22mm]{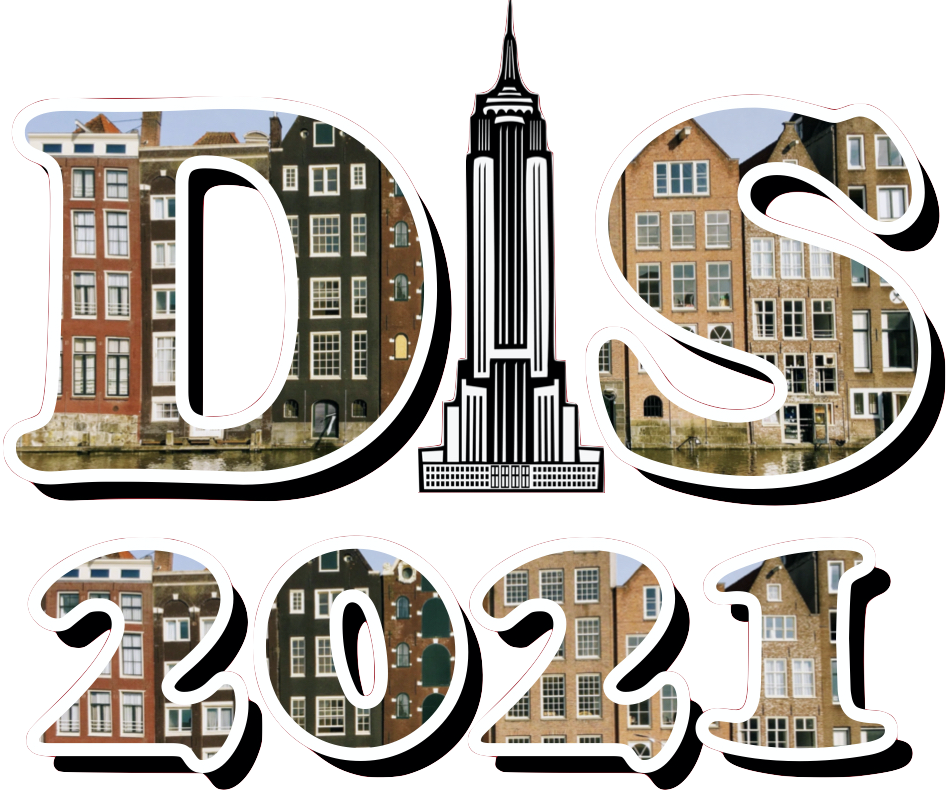}
  \end{minipage}
  &
  \begin{minipage}{0.75\textwidth}
    \begin{center}
    {\it Proceedings for the XXVIII International Workshop\\ on Deep-Inelastic Scattering and
Related Subjects,}\\
    {\it Stony Brook University, New York, USA, 12-16 April 2021} \\
    \doi{10.21468/SciPostPhysProc.?}\\
    \end{center}
  \end{minipage}
\end{tabular}
}
\end{center}

\section*{Abstract}
{\bf

We discuss the sensitivity of the $\gamma\gamma \to \tau^+\tau^-$ process in ultraperipheral Pb+Pb collisions at LHC energies on the anomalous magnetic moment of $\tau$ lepton ($a_\tau$). We derive the corresponding cross sections considering semi-leptonic decays of both leptons in the fiducial volume of ATLAS and CMS detectors. The expected limits on $a_\tau$ with the existing Pb+Pb dataset are found to be better by a factor of two comparing to current best experimental limits and can be further improved by another factor of two at High Luminosity LHC. In addition, our results for tau lepton electric dipole moment, $d_\tau$, can be competitive with the current best limits that were obtained by the Belle experiment.

}


\section{Theoretical framework}
\label{sec:intro}

\begin{figure}[h]
	\centering
	\includegraphics[width=0.5\textwidth]{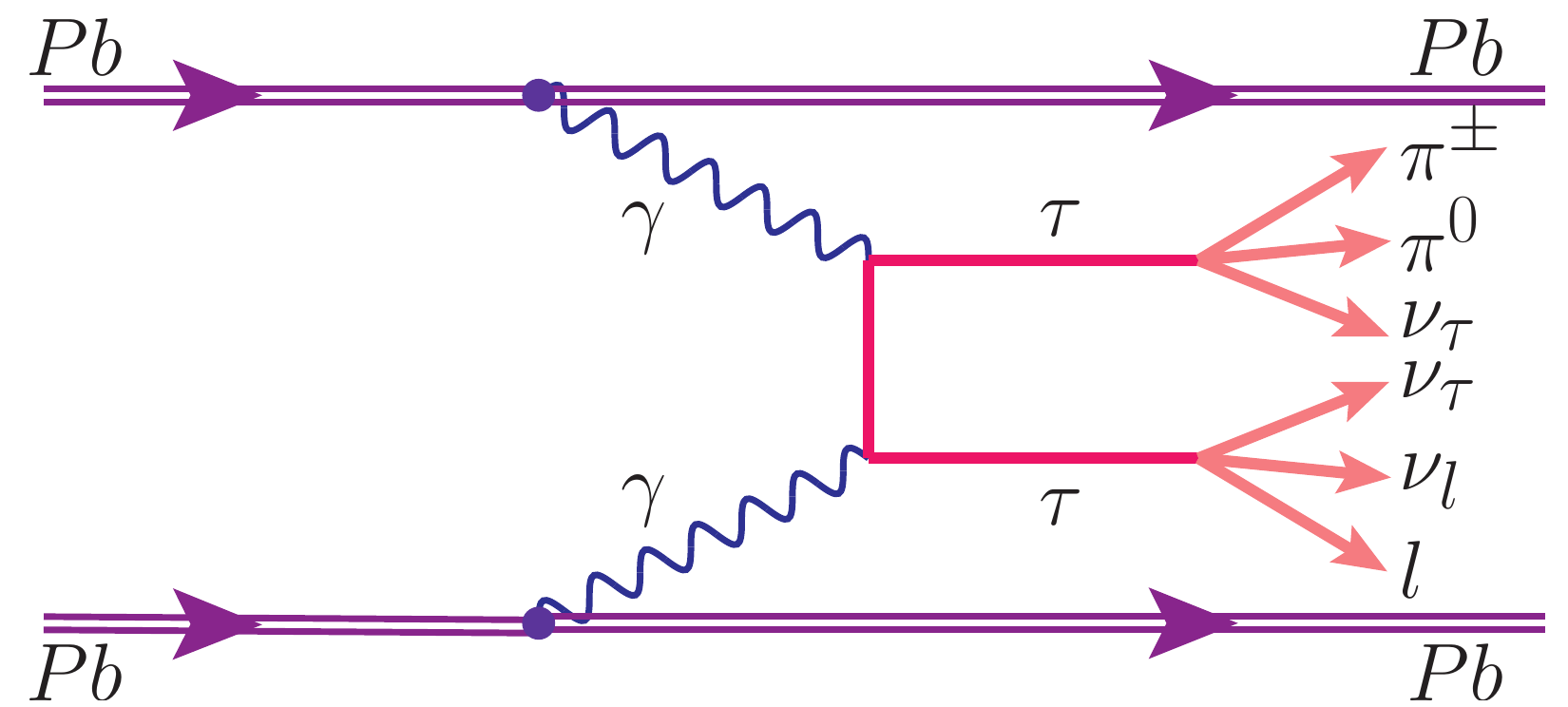}
	\caption{Diagram for the di-taon production in ultraperipheral lead-lead collisions. Main $\tau$ decay channels presented on the graphic, $\tau^\pm\rightarrow \nu_{\tau} +\ell^\pm+ \nu_{\ell}$  ($~\ell=e,~\mu~$) and $\tau^\pm\rightarrow \nu_{\tau}+ \pi^{\pm} + n\pi^{0}~$, give approximately $80\%$ of all $\tau$ decays. }
	\label{fig1}
\end{figure}

The physics of the ultraperipheral collisions (UPC) of heavy ions gives a good opportunity to
study several QED processes~\cite{Baltz:2007kq}. The
Feynman diagram for the Pb+Pb$\rightarrow$Pb+Pb+$\tau^+\tau^-$ process in Fig.~\ref{fig1} includes two $\gamma\tau\tau$ vertices providing an enhanced sensitivity to the anomalous magnetic ($a_{\tau}$) and electric ($d_{\tau}$) moments of the $\tau$ lepton.

Cross section for two-lepton production in heavy-ion collision is the convolution of the elementary cross section for $\gamma\gamma \to \tau^+\tau^-$ and photon fluxes. Due to large charge, ions are surrounded by a strong electromagnetic field. In our approach, photon fluxes depend not only on photon energy but also on the impact parameter \cite{KlusekGawenda:2010kx}. The amplitude for the elementary cross section for the $\gamma \gamma \to \ell^+\ell^-$ reaction in the $t$- and $u$-channel was derived in \cite{Klusek-Gawenda:2017lgt}: 
\begin{eqnarray}
{\mathcal M}
&=&
(-i)\,
\epsilon_{1 \mu}
\epsilon_{2 \nu}
\,\bar{u}(p_{3}) 
\Big(
i\Gamma^{(\gamma \ell\ell)\,\mu}(p_{3},p_{t})
\frac{i(\slash{p}_{t} + m_{\ell})}{t - m_{\ell}^2+i\epsilon}
i\Gamma^{(\gamma \ell\ell)\,\nu}(p_{t'}-p_{4}) \nonumber\\
&&
+
i\Gamma^{(\gamma \ell\ell)\,\nu}(p_{3},p_{u})
\frac{i(\slash{p}_{u} + m_{\ell})}{u - m_{\ell}^2+i\epsilon}
i\Gamma^{(\gamma \ell\ell)\,\mu}(p_{u'}-p_{4}) \Big)
v(p_{4}) \,.
\label{eq:amp_2to2}
\end{eqnarray}
Designating $p'$ and $p$ as momenta of incoming and outgoing lepton and defining $q=p'-p$ as the momentum transfer, a photon-lepton vertex function can be written as:
\begin{equation}
i\Gamma^{(\gamma \ell\ell)}_{\mu}(p',p) = 
-ie\left[ \gamma_{\mu} F_{1}(q^{2})+ \frac{i}{2 m_{\ell}}
\sigma_{\mu \nu} q^{\nu} F_{2}(q^{2}) + \frac{1}{2 m_{\ell}} \gamma^5
\sigma_{\mu \nu} q^{\nu}  F_{3}(q^{2})
\right]
\,,
\label{eq:gamma_lepton_vertex}
\end{equation}
where $\sigma_{\mu \nu} =\dfrac{i}{2}[\gamma_{\mu},\gamma_{\nu}]$, $F_1(q^2)$ and $F_2(q^2)$ are the Dirac and Pauli form factors, $F_3(q^2)$ is the electric dipole form factor.  
The asymptotic values of the form factors, in the $q^2 \rightarrow 0$ limit, are the moments describing the electromagnetic properties of the lepton: $F_1(0) = 1$, $F_2(0) = a_{\ell}$ and $F_3(0) = d_{\ell} \frac{2m_\ell}{e}$.

To study the experimental sensitivity on $a_\tau$ in the $\gamma\gamma\rightarrow\tau^+\tau^-$ processes at the LHC, one has to detect UPC events, which contain two reconstructed $\tau$ leptons and no further activity in the detector. Since $\tau$ lepton is the heaviest lepton with a lifetime of $3 \times 10^{-13}$~s, it decays into lighter leptons ($\tau^\pm\rightarrow \nu_{\tau} +\ell^\pm+ \nu_{\ell},  ~\ell=e,~\mu~$) or hadrons ($\tau^\pm\rightarrow \nu_{\tau}+ \pi^{\pm} + n\pi^{0}~$, $\tau^\pm\rightarrow \nu_{\tau}+ \pi^{\pm}+\pi^{\mp}+\pi^{\pm} + n\pi^{0}$) happens before any direct interaction with the detector material. The reconstruction of tau candidates depends, therefore, on the identification of its unique decay signatures. 
Approximately 80\% of all $\tau$ decays are one charged particle type, and 20\% of them are three-prong decays.

\section{Calculation details}

The nuclear cross section for the Pb+Pb$\rightarrow$Pb+Pb+$\tau^+\tau^-$ process is calculated in the equivalent photon approximation. Next, the \textsc{Pythia8.243} program is used to model $\tau$ decays. 
The QED effect of the final state radiation from outgoing leptons is also simulated by \textsc{Pythia8}. The $\gamma\gamma\rightarrow\tau^+\tau^-$ candidate events are selected by requiring at least one $\tau$ lepton to decay leptonically, as this allows that existing triggering algorithms of the ATLAS or CMS detector can be used~\cite{Aad:2008zzm, Chatrchyan:2008aa}. 
We take into account the events with the limits for the leading electron or muon: $p_{\mathrm{T}}>4$~GeV and $|\eta|<2.5$ . This operation allows an efficient reconstruction and identification by the LHC detectors.

It is worth noting that most produced $\tau$ lepton pairs have relatively low energy (equivalent to low transverse momentum). Therefore, the standard $\tau$ identification tools, developed by the ATLAS and CMS collaborations~\cite{Aad:2015unr, Sirunyan:2018pgf} are not expected to be applicable. We propose, therefore, to categorize the $\gamma\gamma\rightarrow\tau^+\tau^-$ candidate events by their decay mode. All charged-particle tracks from one- or three-prong decays are required to have a transverse momentum of $p_{\mathrm{T}}>0.2$~GeV and a pseudo-rapidity of $|\eta|<2.5$.

The number of events for Pb+Pb$\rightarrow$Pb+Pb+$\tau^+\tau^-$ process \cite{Dyndal:2020yen} for different $a_\tau$ values can be translated into expected sensitivity for limiting $a_{\tau}$.
We treat SM results ($a_{\tau}=0$) as background and the difference between $a_{\tau}=0$ and $a_{\tau}=X$ distributions as a signal.
We use two values of expected systematic uncertainty (5\% and 1\%) and two assumptions on Pb+Pb integrated luminosity (2 nb$^{-1}$ for existing ATLAS/CMS dataset or 20 nb$^{-1}$ for HL-LHC).
The expected significance can be directly transformed into expected 95\% CL limits on $a_{\tau}$. Smaller systematic uncertainty or larger value of luminosity allows predicting a narrower limit on $a_\tau$ \cite{Dyndal:2020yen}.

\section{Main results}

The DELPHI collaboration at LEP2~\cite{Abdallah:2003xd, Cornet:1995pw} obtained the limit: $-0.052 < a_{\tau} < 0.013 ~(95\%~CL)$.
The experimental limits on $a_{\tau}$ were also derived by the L3 and OPAL collaborations in radiative $Z\rightarrow \tau^+\tau^-\gamma$ events at LEP~\cite{Acciarri:1998iv, Ackerstaff:1998mt}, but they are typically weaker by a factor of two comparing to the DELPHI limits.
For comparison, the theoretical Standard Model (SM) value of $a_{\tau}$~\cite{Eidelman:2007sb} is: $a_{\tau}^{\textrm{th}} = 0.00117721 \pm 0.00000005$.

Table~\ref{tab:numbers} contains a summary of the integrated fiducial cross sections at $\sqrt{s_{NN}}=5.02$~TeV for different $a_{\tau}$ values. There is an enumeration of the expected number of reconstructed events in ATLAS or CMS. We assume 80\% reconstruction efficiency within the fiducial region and two values of integrated luminosity ($L_{int}$). The first one corresponds to the existing LHC Pb+Pb dataset: 
$L_{int}=2$~nb$^{-1}$, and the second one relates to expected High Luminosity LHC dataset: $L_{int}=20$~nb$^{-1}$. With the existing Pb+Pb dataset, we expect each experiment to reconstruct about 5000 $\gamma\gamma\rightarrow\tau^+\tau^-$ events ($a_{\tau}=0$). The expected number of reconstructed $\tau$ pairs grows to about 50\,000 at the HL-LHC.

\begin{table}[h]
	\begin{center}
		\begin{tabular}{|l|c|c|c|}
			\hline
			$a_{\tau}$ value & $\sigma_{fid}$ [nb] & Expected events & Expected events \\ 				&	&($L_{int}=2$~nb$^{-1}$, $C=0.8$) & ($L_{int}=20$~nb$^{-1}$, $C=0.8$) \\
			\hline
			$-0.1$ &   4770 & 7650 & 76\,500 \\
			$-0.05$ &  3330 & 5350 & 53\,500 \\
			$-0.02$ &  3060 & 4900 & 49\,000 \\
			\hline
			$0$ (SM) &  3145 & 5050 & 50\,500 \\
			\hline
			$+0.02$ &  3445 & 5500 & 55\,000 \\
			$+0.05$ & 4350 & 6950 & 69\,500 \\
			$+0.1$ & 7225 & 11550 & 115\,500 \\
			\hline
		\end{tabular}
	\end{center}
	\caption{Integrated fiducial cross sections for Pb+Pb$\rightarrow$Pb+Pb+$\tau^+\tau^-$ process for different values of anomalous electromagnetic moments. The expected number of events assuming 80\% selection efficiency and $L_{int}=2$~nb$^{-1}$ or $L_{int}=20$~nb$^{-1}$ are also shown.}
	\label{tab:numbers}
\end{table}

The number of events from Table~\ref{tab:numbers} can be translated into expected sensitivity for probing $a_{\tau}$.
We use the \textsc{RooFit} toolkit for the statistical analysis of the results.
We perform fits to $R_{\ell}(p_{\mathrm{T}}^{lead~lepton})$ distribution by treating SM results ($a_{\tau}=0$) as background and the difference between $a_{\tau}=0$ and $a_{\tau}=X$ distributions as a signal.
The procedure exploits both normalization and $p_{\mathrm{T}}^{lead~lepton}$ shape differences, providing extra sensitivity on $a_{\tau}$ measurement.
We use two values of expected systematic uncertainty (5\% and 1\%) and two assumptions on Pb+Pb integrated luminosity (2 nb$^{-1}$ for existing ATLAS/CMS dataset or 20 nb$^{-1}$ for HL-LHC).

\begin{figure}[h]
	\centering
	(a)\includegraphics[width=0.45\textwidth]{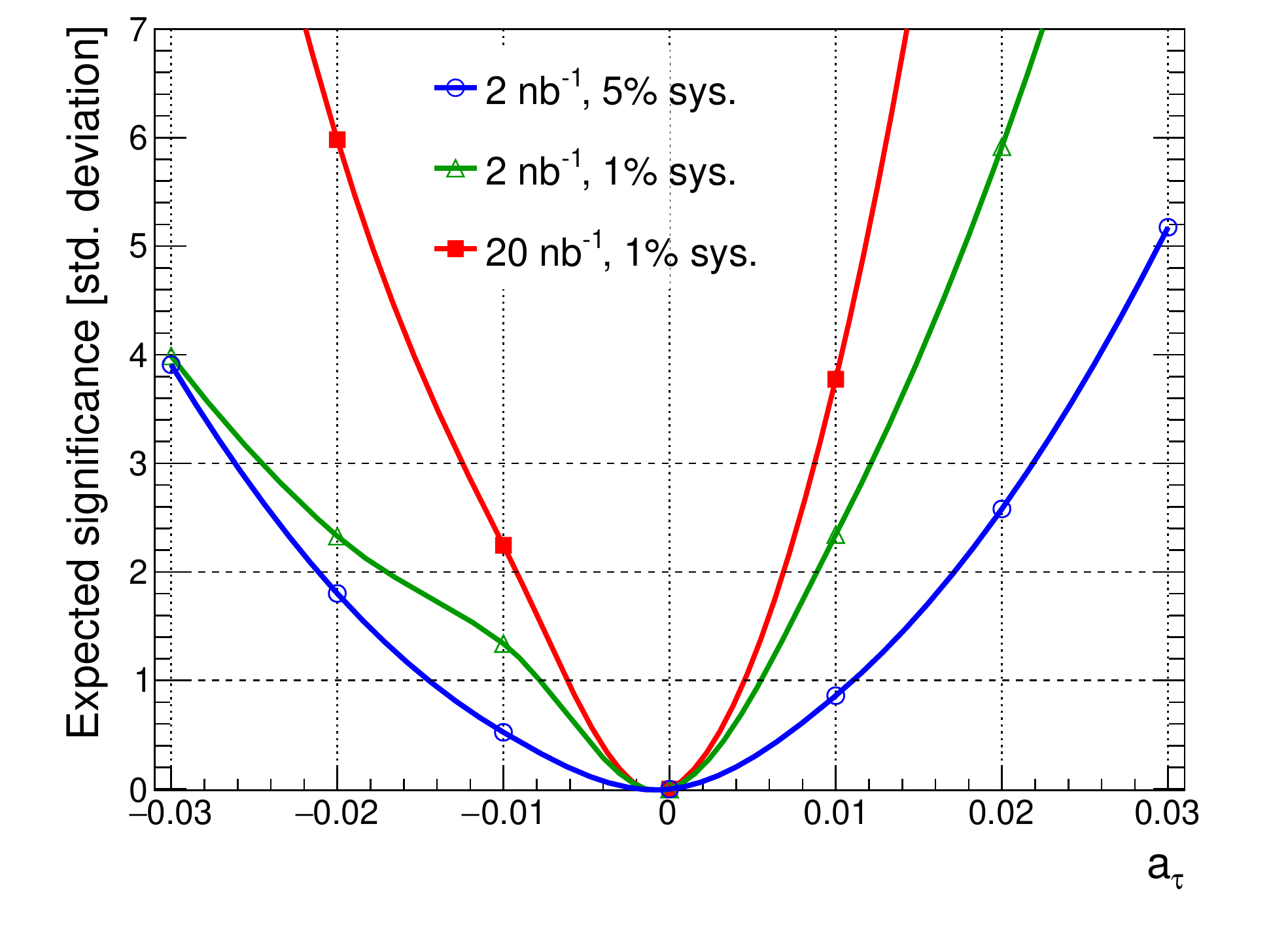}
	(b)\includegraphics[width=0.45\textwidth]{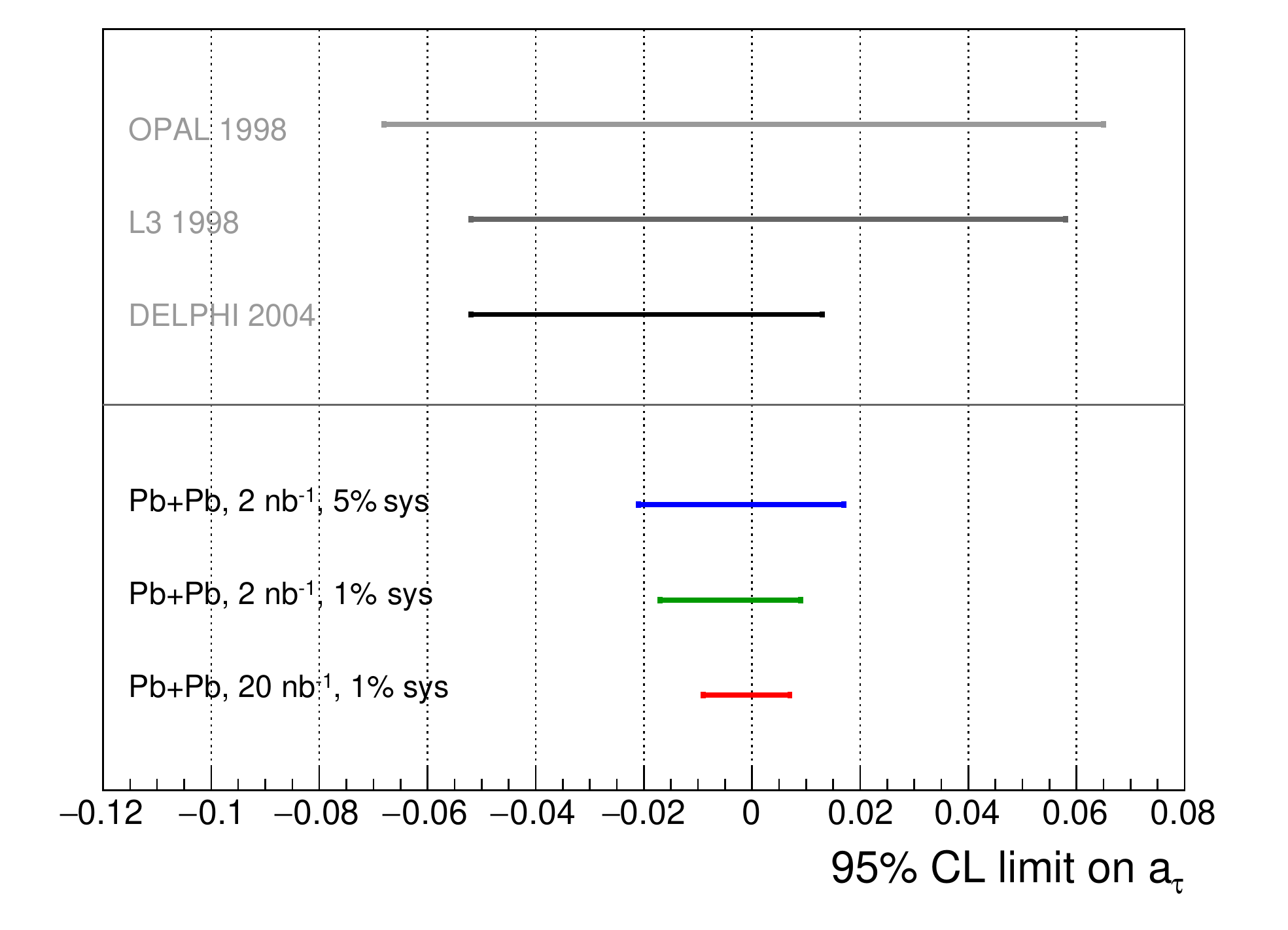}
	\caption{(a) Expected signal significance as a function of anomalous $\tau$ moment for different values of Pb+Pb integrated luminosity and total systematic uncertainty. \\
	(b) Expected 95\% CL limits on $a_{\tau}$ measurement for different values of the Pb+Pb integrated luminosity and total systematic uncertainty. Comparison is also made to the existing limits from OPAL~\cite{Ackerstaff:1998mt}, L3~\cite{Acciarri:1998iv} and DELPHI~\cite{Abdallah:2003xd} experiments at LEP.}
	\label{fig2}
\end{figure}

Figure~\ref{fig2}(a) shows the expected signal significance as a function of $a_{\tau}$. The observed asymmetry for positive and negative $a_{\tau}$ values reflects the destructive interference between SM and the anomalous $\tau$ coupling.
The expected significance can be directly transformed into expected 95\% CL limits on $a_{\tau}$, shown in Fig.~\ref{fig2}(b). Assuming 2 nb$^{-1}$ of integrated Pb+Pb luminosity and 5\% systematic uncertainty, the expected limits are $-0.021<a_{\tau}<0.017$, approximately two times better than the DELPHI limits~\cite{Abdallah:2003xd}.
By collecting more data (20 nb$^{-1}$) and with improved systematic uncertainties, these limits can be further improved by another factor of two.
The expected results by studying ultraperipheral collisions at the LHC can significantly improve the existing limits on $a_{\tau}$.

\section{Conclusion}

Here we presented a prediction on the cross section of the $\gamma \gamma \to \tau^+ \tau^-$ process and its dependence on anomalous electromagnetic couplings of the tau lepton in ultraperipheral Pb+Pb collisions at the LHC.  
We also investigated the expected sensitivity on $a_{\tau}$ and $d_{\tau}$, assuming standard LHC detectors using the currently available and future datasets. We propose to use cross section ratios of the $\gamma \gamma \to \tau^+ \tau^-$  and $\gamma \gamma \to e^+ e^- (\mu^+\mu^-)$ processes to probe $a_{\tau}$, as several systematic uncertainties cancel and the experimental knowledge of $a_e$ and $a_\mu$ is several orders of magnitude more precise than $a_\tau$ itself.

Our studies suggest that the currently available datasets of the LHC experiments are already sufficient to improve the sensitivity on $a_\tau$ by a factor of two. Hence, we consider experimental analysis as highly interesting and worthwhile to be done in the future.
Future Belle-II experiment should give much better constraints on
$|a_{\tau}|<1.75\cdot10^{-5}$ and $|d_{\tau}|<2.04\cdot10^{-19}$~$e\cdot\textrm{cm}$ 
\cite{Chen:2018cxt}.

We have also studied the sensitivity on tau lepton electric dipole moment, $d_{\tau}$.
Our expected 95\% CL sensitivity on $|d_{\tau}|$ assuming $a_{\tau}=0$ is: $|d_{\tau}|<6.3~(4.4)\cdot10^{-17}$ $e\cdot\textrm{cm}$ at the LHC with 5\% (1\%) systematic uncertainty and $|d_{\tau}|<3.5\cdot10^{-17}$ $e\cdot\textrm{cm}$ at HL-LHC (1\% systematic uncertainty). 
For comparison, the current best limits are measured by Belle experiment~\cite{Inami:2002ah}: $-2.2 < Re(d_{\tau}) < 4.5 ~(10^{-17}~e\cdot\textrm{cm})$ and $-2.5 < Im(d_{\tau}) < 0.8 ~(10^{-17}~e\cdot\textrm{cm})$.
Our projected results on $d_{\tau}$ can be therefore competitive with the Belle limits.
\section*{Acknowledgements}

\paragraph{Funding information}
This study was partially supported by the Polish National Science Center grant
UMO-2018/31/B/ST2/03537 and by the Center for Innovation and Transfer of Natural Sciences and Engineering Knowledge in Rzeszów. 
The project is co-financed by the Polish National Agency for Academic Exchange within Polish Returns Programme, Grant No. PPN/PPO/2020/1/ 00002/U/00001.


\bibliographystyle{SciPost_bibstyle}
\bibliography{biblio}




\nolinenumbers

\end{document}